\documentclass[preprint]{aastex631}
\shorttitle{Al-coated sCMOS}
\shortauthors{Wu et al.}
\graphicspath{{./}{figure/}}

\begin{document}


\title{An Aluminum-coated sCMOS sensor for X-Ray Astronomy}

\author{Qinyu Wu}
\affiliation{National Astronomical Observatories, Chinese Academy of Sciences \\
20A Datun Road, Chaoyang District \\
Beijing 100101, China}
\affiliation{School of Astronomy and Space Science, University of Chinese Academy of Sciences \\
19A Yuquan Road, Shĳingshan District \\
Beijing 100049, China}

\author{Zhixing Ling}
\affiliation{National Astronomical Observatories, Chinese Academy of Sciences \\
20A Datun Road, Chaoyang District \\
Beijing 100101, China}
\affiliation{School of Astronomy and Space Science, University of Chinese Academy of Sciences \\
19A Yuquan Road, Shĳingshan District \\
Beijing 100049, China}
\correspondingauthor{Zhixing Ling}
\email{lingzhixing@nao.cas.cn}


\author{Chen Zhang}
\affiliation{National Astronomical Observatories, Chinese Academy of Sciences \\
20A Datun Road, Chaoyang District \\
Beijing 100101, China}
\affiliation{School of Astronomy and Space Science, University of Chinese Academy of Sciences \\
19A Yuquan Road, Shĳingshan District \\
Beijing 100049, China}

\author{Shuang-Nan Zhang}
\affiliation{National Astronomical Observatories, Chinese Academy of Sciences \\
20A Datun Road, Chaoyang District \\
Beijing 100101, China}
\affiliation{School of Astronomy and Space Science, University of Chinese Academy of Sciences \\
19A Yuquan Road, Shĳingshan District \\
Beijing 100049, China}
\affiliation{Institute of High Energy Physics, Chinese Academy of Sciences \\
19B Yuquan Road, Shĳingshan District \\
Beijing 100049, China}

\author{Weimin Yuan}
\affiliation{National Astronomical Observatories, Chinese Academy of Sciences \\
20A Datun Road, Chaoyang District \\
Beijing 100101, China}
\affiliation{School of Astronomy and Space Science, University of Chinese Academy of Sciences \\
19A Yuquan Road, Shĳingshan District \\
Beijing 100049, China}



\begin{abstract}

In recent years, tremendous progress has been made on scientific Complementary Metal Oxide Semiconductor (sCMOS) sensors, making them a promising device for future space X-ray missions. We have customized a large-format sCMOS sensor, G\-1516\-BI, dedicated for X-ray applications. In this work, a 200 nm thick aluminum layer is successfully sputtered on the surface of this sensor. This Al-coated sensor, named EP4K, shows consistent performance with the uncoated version. The readout noise of the EP4K sensor is around 2.5 $\rm{e^{-}}$ and the dark current is less than 0.01 $\rm{e^-/pixel/s}$ at $\rm{-30 ^{\circ}\!C}$. The maximum frame rate is 20 Hz in the current design. The ratio of single pixel events of the sensor is $45.0\%$. The energy resolution can reach 153.2 eV at 4.51 keV and 174.2 eV at 5.90 keV at $\rm{-30 ^{\circ}\!C}$. The optical transmittance of the aluminum layer is approximately $10^{-8}$ to $10^{-10}$ for optical lights from 365 to 880 nm, corresponding to an effective aluminum thickness of around 140 to 160 nm. The good X-ray performance and low optical transmittance of this Al-coated sCMOS sensor make it a good choice for space X-ray missions. The Lobster Eye Imager for Astronomy (LEIA), which has been working in orbit for about one year, is equipped with four pieces of EP4K sensors. Furthermore, 48 pieces of EP4K sensors are used on the Wide-field X-ray Telescope (WXT) on the Einstein Probe (EP) satellite, which will be launched at the end of 2023.

\end{abstract}



\section{Introduction}
\label{sect:intro}
Silicon image sensors, including charge-coupled devices (CCDs) and Complementary Metal Oxide Semiconductor (CMOS) sensors, are widely used in space missions, especially for X-ray telescopes. However, these sensors are sensitive to not only X-rays but also optical and UV light. Therefore, to maximize their X-ray detection capabilities, optical blocking filters (OBFs) must be inserted in the light path of the X-rays. A free-standing aluminum-coated polymer filter is the main choice for most X-ray telescopes in the past, including Chandra ACIS \citep{Chandra_ACIS_CCD_calibration_1998, Chandra_ACIS_filter_1996, Chandra_ACIS_filter_calibration_1996}, XMM-Newton EPIC \citep{XMMNewton_EPIC_filter_free}, Suzaku XIS \citep{Suzaku_XIS_CCD_good_FWHM} and Swift XRT \citep{Swift_XRT_2005}. Some attempts have also been made by coating a filter directly onto a sensor over the last few years. Five cameras on the eROSITA telescopes used Al-coated pn-CCDs as focal plane detectors \citep{eRosita_CCD_Meidinger_2021_good_FWHM, eRosita_CCD_PhD_thesis}. The CCD detectors of the MAXI SSC instrument were also coated with 200 nm aluminum directly on the surfaces of the sensors \citep{MAXI_SSC_2011}. 

All of the aforementioned experiments have employed CCDs as focal plane detectors, showing the dominance of CCDs in the field of soft X-ray detection over the past several decades. However, in recent years, the performance of sCMOS sensors has been significantly improved, boasting faster readout speeds, superior irradiation resistance, and more flexible operating temperatures compared to traditional CCDs. Therefore, several X-ray space missions, such as LEIA \citep{LEIA_Zhang_APJL_2022, LEIA_RAA}, EP \citep{EP_2018, EP_2022}, and THESEUS (concept) \citep{Heymes_2020_CIS221_good_FWHM, CIS221_2022_good_FWHM_conference, CIS221_2022_good_FWHM}, used or will use sCMOS sensors as focal plane detectors. 

Our laboratory has studied the X-ray performances of several sCMOS sensors \citep{Wang_2019_JINST_G400, Ling_2021_crosstalk_G400}. We customized a large-format back-illuminated sCMOS sensor, G\-1516\-BI, in 2019, which has an array of $4096\times4096$ pixels with a pixel size of $\rm{15\ \mu m}$ \citep{Wu_PASP_2022}. It has a fully depleted epitaxial layer of $\rm{10\ \mu m}$. The frame rate is 20 fps in the current design. At $\rm{-30 ^{\circ}\!C}$, the dark current is lower than 0.05 $\rm{e^-}$/pixel/s and the readout noise is lower than 5 $\rm{e^-}$. The energy resolution is about 180 eV at 5.9 keV at room temperature, and can be improved to 124.6 eV at 4.5 keV by making a pixel level gain correction \citep{Wu_2023_PASP}. The image lag of this sensor is less than 0.5 $\rm{e^-}$ \citep{Wu_2023_NIMA}. G\-1516\-BI is not equipped with any OBF.

We have evaluated the influence of different Al coating layers on a medium-format sCMOS sensor \citep{Wang_2022_JINST_G400Al}. The optical transmittance at 660 and 850 nm can reach a level of about $10^{-9}$ for a 200 nm aluminum layer and about $10^{-4}$ for a 100 nm aluminum layer. Based on these results, a 200 nm thick aluminum layer is uniformly sputtered on the entrance window of the large-format sensor, G\-1516\-BI. This Al-coated sensor, named EP4K, was successfully fabricated in 2021. The design and basic properties of this Al-coated sCMOS sensor are given in Section~\ref{sect:basic_propt}; the optical transmittance results are shown and discussed in Section~\ref{sect:transmittance}; and the conclusions are summarized in Section~\ref{sect:concls}.

\section{Design and Basic Properties}
\label{sect:basic_propt}
\begin{figure}
\begin{center}
\begin{tabular}{c}
\includegraphics[width=0.9\textwidth]{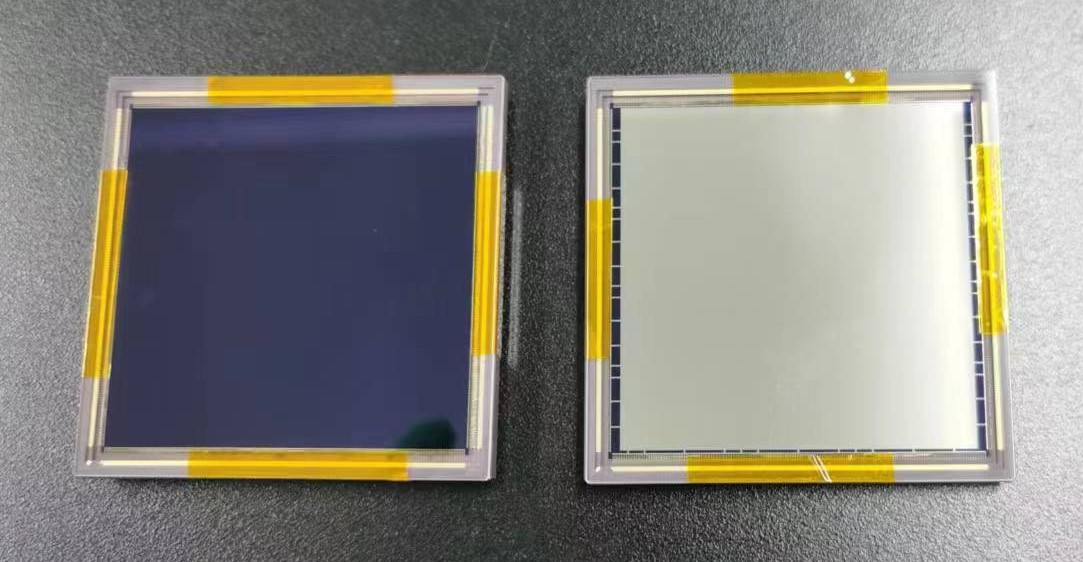}
\end{tabular}
\end{center}
\caption{The uncoated sensor (left) and the Al-coated EP4K sensor (right). The photosensitive area of the two sensors is 6 cm $\times$ 6 cm, which contains an array of $4096\times4096$ pixels with a pixel size of $\rm{15\ \mu m}$. The Al-coated layer on the EP4K sensor is grounded through the aluminum wires at its edges.}
\label{fig:chip_illustration}
\end{figure}

\begin{figure}
\begin{center}
\begin{tabular}{c}
\includegraphics[width=0.6\textwidth]{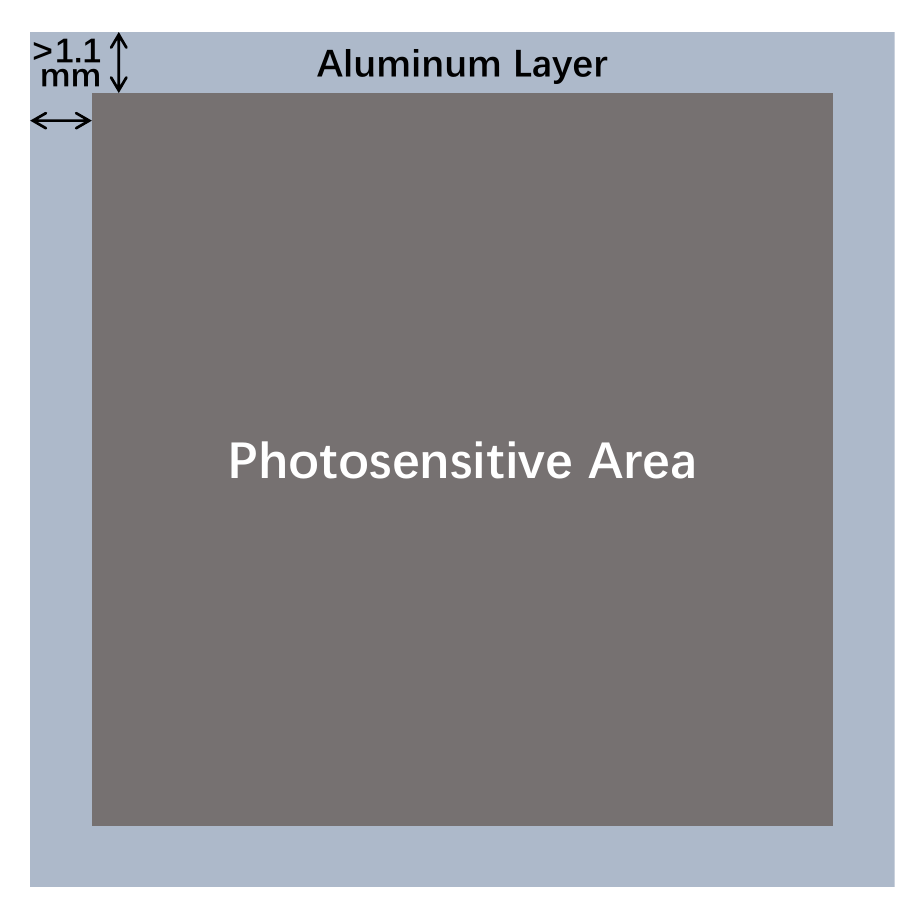}
\end{tabular}
\end{center}
\caption 
{ \label{fig:coating_design}
Illustration of the aluminum coating on the surface of the EP4K sCMOS sensor. The Al film is extended by more than 1.1 mm out of the four edges of the photosensitive area of the sensor to block the light penetrating from the edges of the silicon die.} 
\end{figure}

The uncoated G\-1516\-BI sensor (left) and the Al-coated EP4K sensor (right) are shown in Fig.~\ref{fig:chip_illustration}. The EP4K sensor has a 200 nm thick aluminum film coated on its surface. As shown in Fig.~\ref{fig:coating_design}, to reduce the penetration of light from the edges of the silicon die, the covered area of the aluminum film is extended by more than 1.1 mm at the four edges of the sensitive area of the sensor. To investigate the basic properties of this Al-coated sensor and to check the influence of the Al coating, the uncoated G\-1516\-BI sensor is used as a reference. Dark exposure experiments are performed on these two sensors at different temperatures to obtain the readout noise and the dark current level. Then, the X-ray responses are tested with several X-ray sources.

\subsection{Readout Noise}
\label{subsect:readout_noise}
\begin{figure}
\begin{center}
\begin{tabular}{c}
\includegraphics[width=0.75\textwidth]{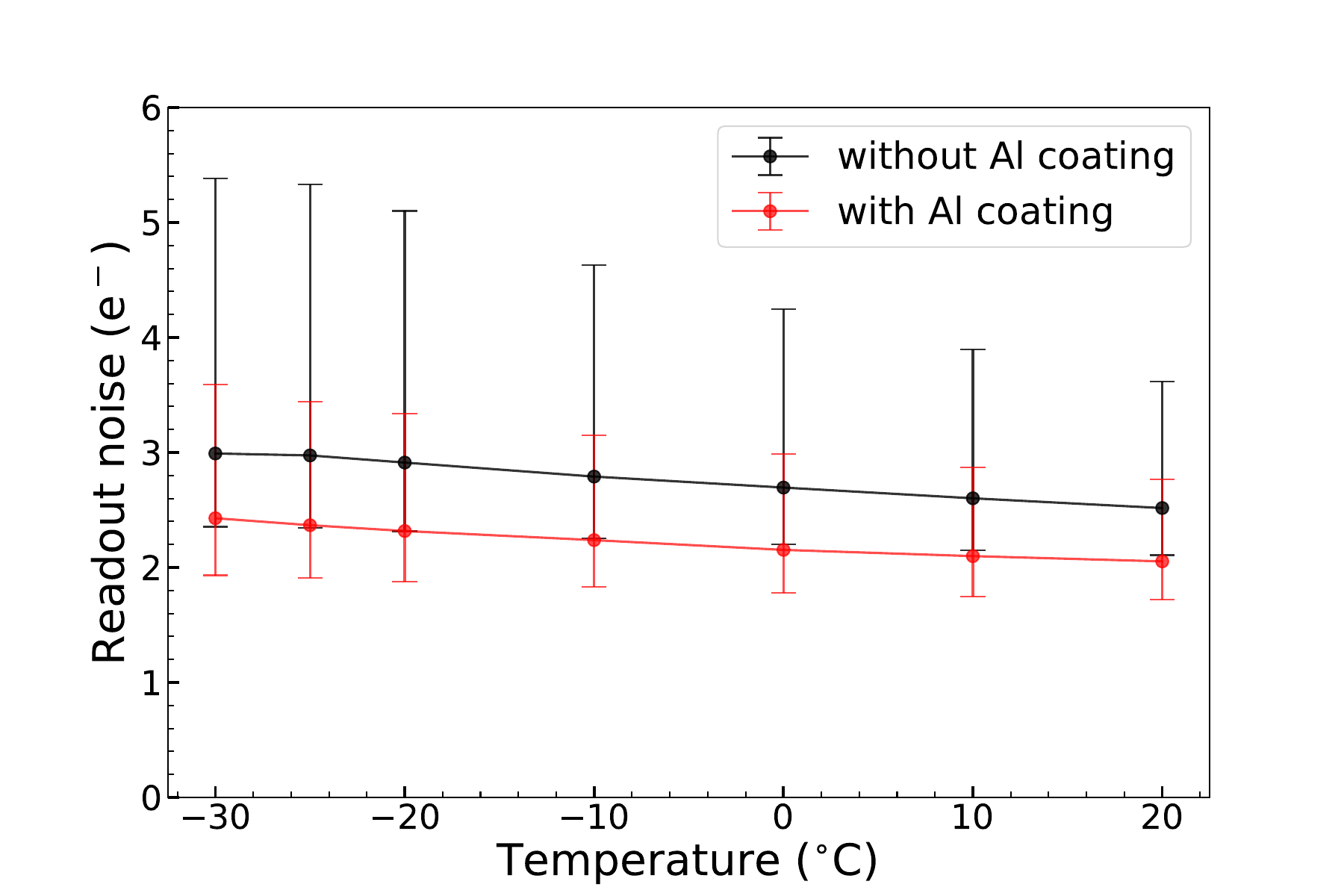}
\end{tabular}
\end{center}
\caption 
{ \label{fig:noise_temp}
The readout noise at different temperatures, for the Al-coated sensor (red) and the uncoated sensor (black), respectively. The readout noise is measured as the standard deviation among fifty frames in a dark environment. The error bars represent the range between the 16\% percentile and the 84\% percentile of the readout noise among pixels.} 
\end{figure}

The readout noise is measured at different temperatures, ranging from $\rm{-30 ^{\circ}\!C}$ to $\rm{20 ^{\circ}\!C}$. At each temperature, fifty frames are recorded in a dark environment with the shortest integration time of $\rm{13.92\ \mu s}$. The standard deviation among these frames gives a readout noise map of the sensor. The median value of this noise map among all pixels is chosen to represent the readout noise level of the whole sensor. As shown in Fig.~\ref{fig:noise_temp}, the readout noise levels of the Al-coated sensor and the uncoated sensor are similar: both are around 3 $\rm{e^{-}}$ and decrease with increasing temperature. The error bars in Fig.~\ref{fig:noise_temp} represent the variances in the readout noise among pixels, showing that the uncoated sensor has a larger pixel-to-pixel variation than the Al-coated sensor. This is likely due to individual variations between the two sensors, rather than a systematic difference between the two types of sensors.

\subsection{Dark Current}
\label{subsect:dark_current}
\begin{figure}
\begin{center}
\begin{tabular}{c}
\includegraphics[width=0.75\textwidth]{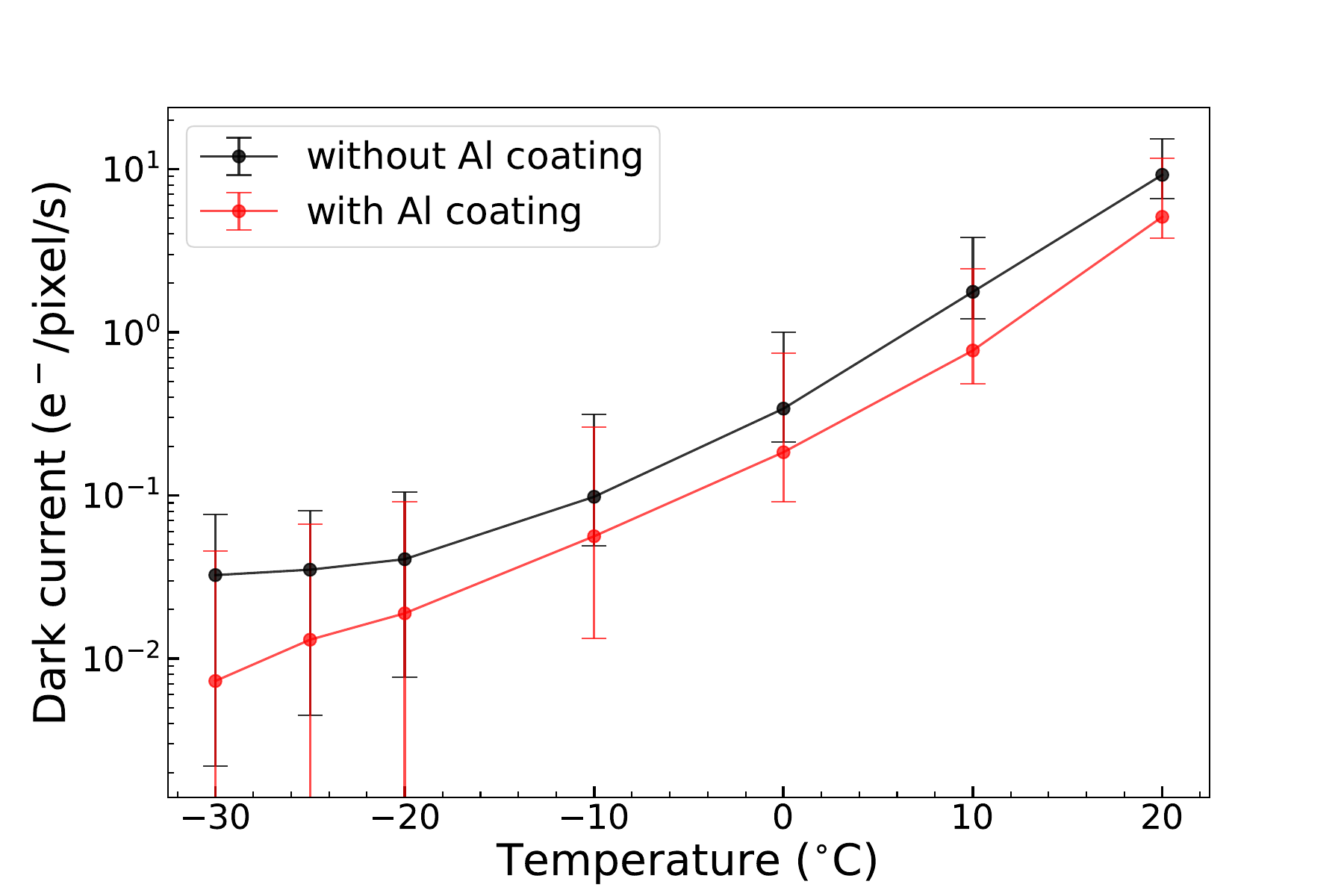}
\end{tabular}
\end{center}
\caption 
{ \label{fig:dc_temp}
The dark current at different temperatures, for the Al-coated sensor (red) and the uncoated sensor (black), respectively. The error bars represent the range between the 16\% percentile and the 84\% percentile of the dark current among pixels.} 
\end{figure}

At different temperatures, the dark current of each pixel can be calculated as the slope of the linear correlation between the signal values and the integration times, ranging from $\rm{13.92\ \mu s}$ to 100 s; the median value of the dark current map represents the dark current level of the whole sensor. Fig.~\ref{fig:dc_temp} shows that the dark current levels of the two sensors are similar and follow the same trends as the temperature increases. The dark currents of both sensors are lower than $0.05\ \rm{e^{-}/pixel/s}$ at $\rm{-30 ^{\circ}\!C}$, which can have negligible influence on performance. The dark current of the uncoated sensor is slightly higher than that of the Al-coated sensor, which is also due to the device-to-device variations.

\subsection{X-Ray Response}
\label{subsect:xray_response}
The X-ray responses of the Al-coated sensor and the uncoated sensor are tested with an $^{55}\!\rm{Fe}$ source and an X-ray tube with different targets, including Si and Ti, respectively. The two sensors are exposed to each type of X-ray source at $\rm{-30 ^{\circ}\!C}$. For each of the frames taken in the X-ray exposures, events are searched over the image and recorded in real time by the camera, as described in detail in \citet{Wu_PASP_2022} and \citet{wang_2022_camera}. In one image, if the value of a pixel, or Digital Number (DN), is above a preset threshold, ${T_{\rm event}}$, and is the maximum among its adjacent $3\times3$ pixels, then this region is taken as an event. A single pixel event, or Grade 0 event (see Section 3 of \citet{Wu_PASP_2022} for the definition of the grade of an event), is defined as that no other pixels than the center one are above the split threshold, ${T_{\rm split}}$, which is set to a value about ten times of the readout noise level. Two types of spectra are built: GAtotal and G0center spectrum. The GAtotal spectrum is extracted from events of all kinds of grade, and the total charge in the $3\times3$ region of each event is used. In contrast, the G0center spectrum is built from the center pixel's charge of each single pixel event. 

\begin{figure}[htbp]
\centering
\resizebox{\hsize}{!}{
\includegraphics[width=0.45\textwidth]{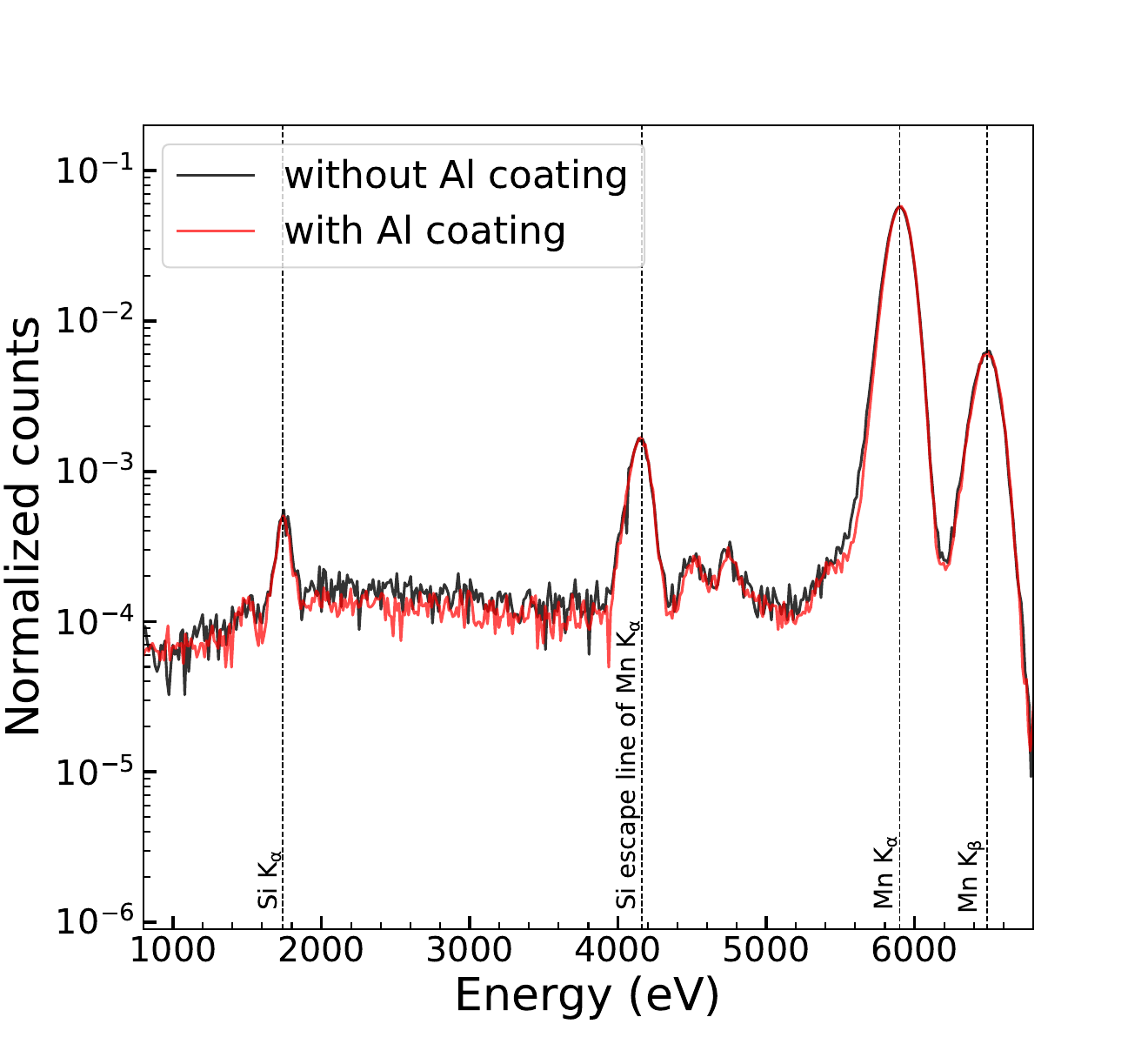}
\includegraphics[width=0.45\textwidth]{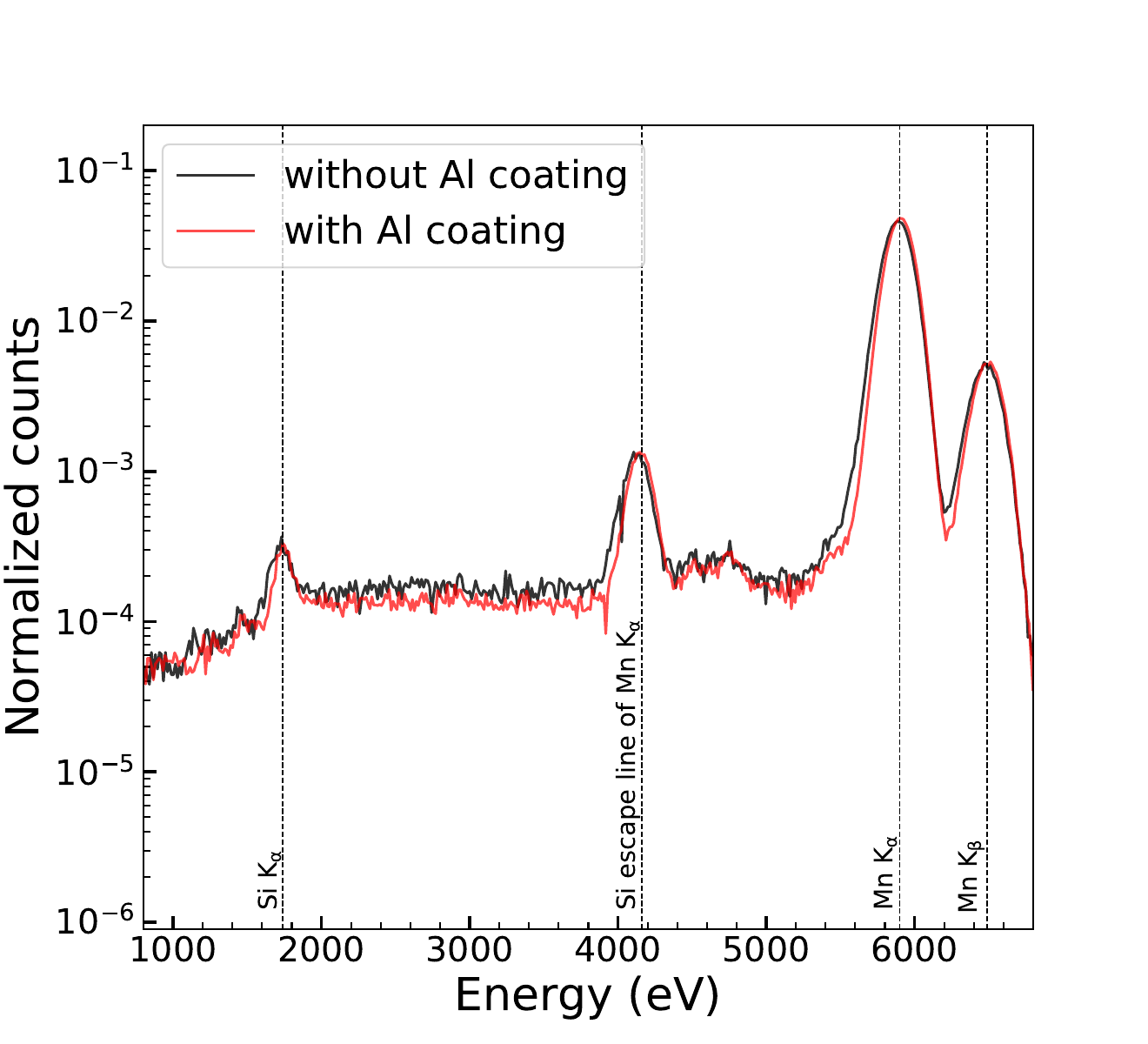}}
\caption{The normalized G0center spectra (left) and GAtotal spectra (right) of the Al-coated sensor (red) and the uncoated sensor (black) irradiated by an $^{55}\!\rm{Fe}$ source. Four characteristic lines can be identified: Si $\rm{K_\alpha}$ (1.74 keV), Si escape peak of Mn $\rm{K_\alpha}$ (4.16 keV), Mn $\rm{K_\alpha}$ (5.90 keV), and Mn $\rm{K_\beta}$ (6.49 keV).}
\label{fig:spec_compare}
\end{figure}

Fig.~\ref{fig:spec_compare} gives the G0center and GAtotal spectra of the two sensors irradiated by an $^{55}\!\rm{Fe}$ source, where no significant differences can be seen. Then the characteristic lines from not only the $^{55}\!\rm{Fe}$ source but also the Si and Ti targets are used to study the conversion gains and energy resolutions of the two sensors. To determine the peak location and the Full Width Half Maximum (FWHM), each of the peaks is fitted with a simple Gaussian function. The conversion gain can then be obtained by a linear fitting on the relationship between the locations of these peaks in the spectrum and their corresponding energies. Table~\ref{table:x_ray_prop} shows the X-ray performances of the Al-coated and the uncoated sensors with the results from the G0center and GAtotal spectra. Compared to the uncoated sensor, the Al-coated version has similar single pixel event ratios, conversion gains, and a little bit better energy resolutions. These differences may be related to the manufacturing process of the two individual sensors. The better energy resolutions of the Al-coated sensor are partly due to its lower readout noise level. The possible difference in gain uniformity may also contribute to the resolution difference.

\begin{table*}[htbp]
\centering
\resizebox{\textwidth}{!}{
\begin{tabular}{| l | l | l | l | l |}
\hline
Properties & \multicolumn{2}{| c |}{Al-coated sensor} & \multicolumn{2}{| c |}{Uncoated sensor}\\ \hline
& G0center & GAtotal & G0center & GAtotal \\ \hline
Ratio of single pixel events (\%) & \multicolumn{2}{| c |}{$45.0 \pm 0.2 $} & \multicolumn{2}{| c |}{$43.3 \pm 0.3 $}\\ \hline
Conversion gain (eV/DN) & $6.50\pm0.01$ & $6.53\pm0.01$ & $6.69\pm0.01$ & $6.76\pm0.01$ \\ \hline
FWHM at Si $\rm{K_\alpha}$, 1.74 keV (eV) & $98.8\pm1.9$ & $138.3\pm0.8$ & $103.5\pm2.2$ & $152.7\pm0.8$ \\ \hline
FWHM at Ti $\rm{K_\alpha}$, 4.51 keV (eV) & $153.2\pm1.2$ & $186.0\pm0.5$ & $156.6\pm1.3$ & $200.3\pm0.7$ \\ \hline
FWHM at Mn $\rm{K_\alpha}$, 5.90 keV (eV) & $174.2\pm1.1$ & $209.7\pm0.7$ & $179.2\pm1.3$ & $224.5\pm0.9$ \\ \hline
\end{tabular}}
\caption{The X-ray performances of the Al-coated and the uncoated sCMOS sensor at $\rm{-30 ^{\circ}\!C}$.}
\label{table:x_ray_prop}
\end{table*}

\section{Optical Transmittance}
\label{sect:transmittance}
The optical transmittance of the Al-coated EP4K sensor is measured with LED panel sources of different wavelengths, ranging from 365 to 880 nm. In each experiment, the LED source of a given wavelength is placed 40 cm away facing the sensor so that the light can incident vertically into the sensor. As the input power of the LED sources can be adjusted, a photodiode power sensor is placed directly next to the sCMOS sensor to measure the light intensity $P$. An averaged exposure map, $M$, is calculated from 30 frames recorded by the sensor with a given integration time $t$. To calculate the transmittance of the Al layer at a given wavelength, we perform the above measurements on the Al-coated EP4K and the uncoated G\-1516\-BI sensors, respectively. The transmittance map is then given by:
\begin{equation}
\label{eq:transmittance}
    T = \frac{M_{\rm Al}}{t_{\rm Al} \times P_{\rm Al}} \div \frac{M_{\rm NoAl}}{t_{\rm NoAl} \times P_{\rm NoAl}}.
\end{equation}
Since the Analogue to Digital Converters (ADCs) of these sensors have 12 bits, the dynamic range of the sensor is around 4000, which is much larger than the transmittance around $10^{-9}$. Therefore, the integration times of the two sensors are set as $t_{\rm Al}=1200\ \rm{ms}$ and $t_{\rm NoAl}=5\ \rm{ms}$. The ratio between the power of the LED, $\frac{P_{\rm Al}}{P_{\rm NoAl}}$, is set to around $10^4$ to $10^5$ in each pair of measurements.

\begin{figure}
\begin{center}
\begin{tabular}{c}
\includegraphics[width=0.8\textwidth]{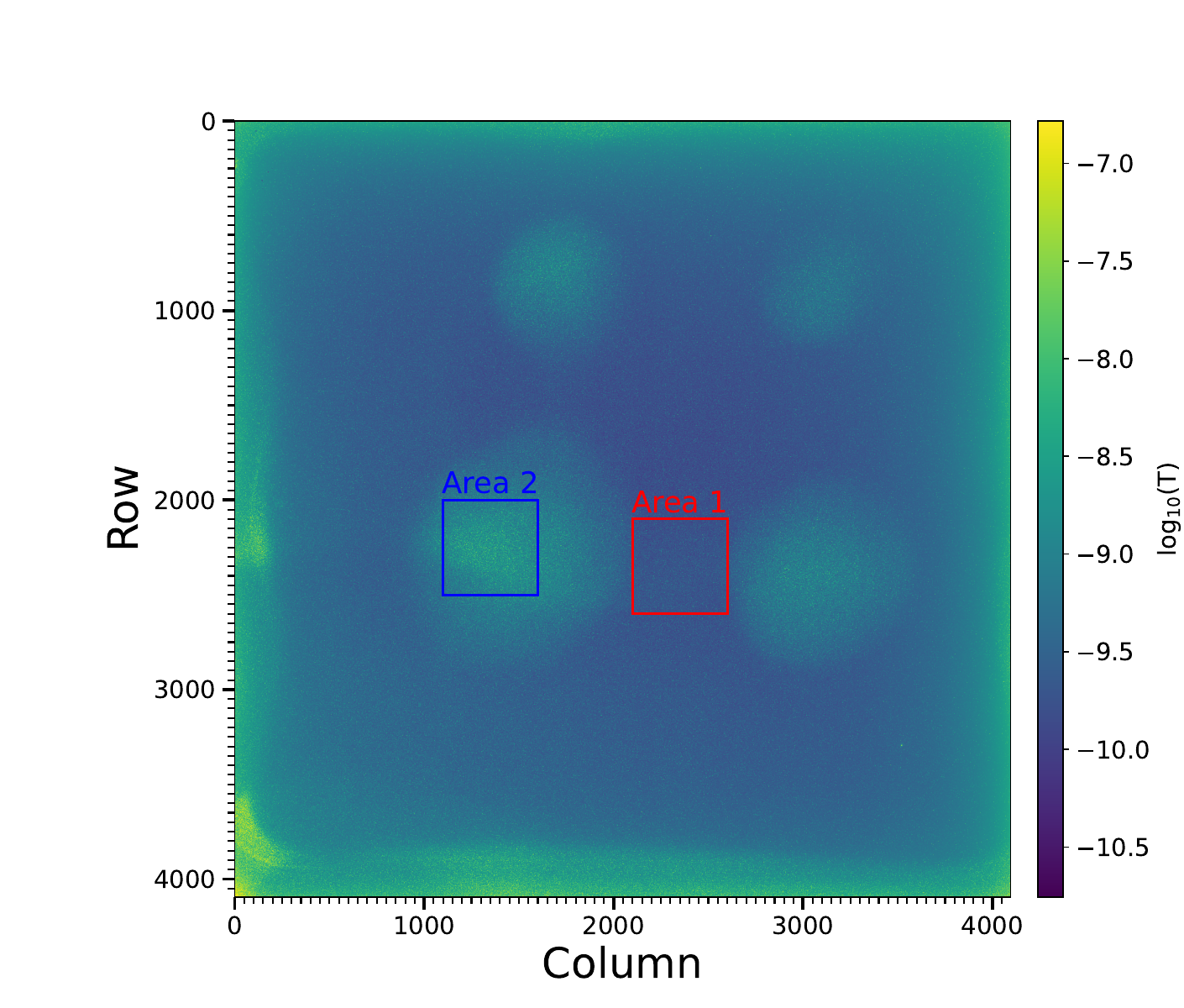}
\end{tabular}
\end{center}
\caption 
{ \label{fig:trans_map}
The transmittance map of the Al-coated sCMOS sensor at a wavelength of 570 nm. Two typical areas are selected for further study: Area 1 (red) is outside of the high-transmittance pattern, and Area 2 (blue) is inside the pattern.} 
\end{figure}

Fig.~\ref{fig:trans_map} shows the transmittance map of the Al-coated sensor at a wavelength of 570 nm. The median transmittance of the whole sensor is $4.2\times10^{-10}$, and the transmittance in most areas are lower than $10^{-9}$. Although it has improved a lot compared to \citet{Wang_2022_JINST_G400Al}, light leakage from the four edges of the sensor can also be seen, giving a median transmittance of $3.6\times10^{-9}$ in the range of 50 pixels from the edge of the sensor. Some high-transmittance patterns are observed in the central region of the map, which may be related to the imperfect sputter process. Therefore, we have chosen Area 1, which is outside the pattern, and Area 2, which is inside the pattern, for further study. At 570 nm, Area 1 has a median transmittance of $1.8\times10^{-10}$, which is around 5 times lower than that of Area 2 with a median of $9.6\times10^{-10}$. From Fig.~\ref{fig:trans_map}, some pinholes with transmittance larger than $10^{-7}$ may exist, which will be studied in detail in the future.

\begin{figure}
\begin{center}
\begin{tabular}{c}
\includegraphics[width=0.75\textwidth]{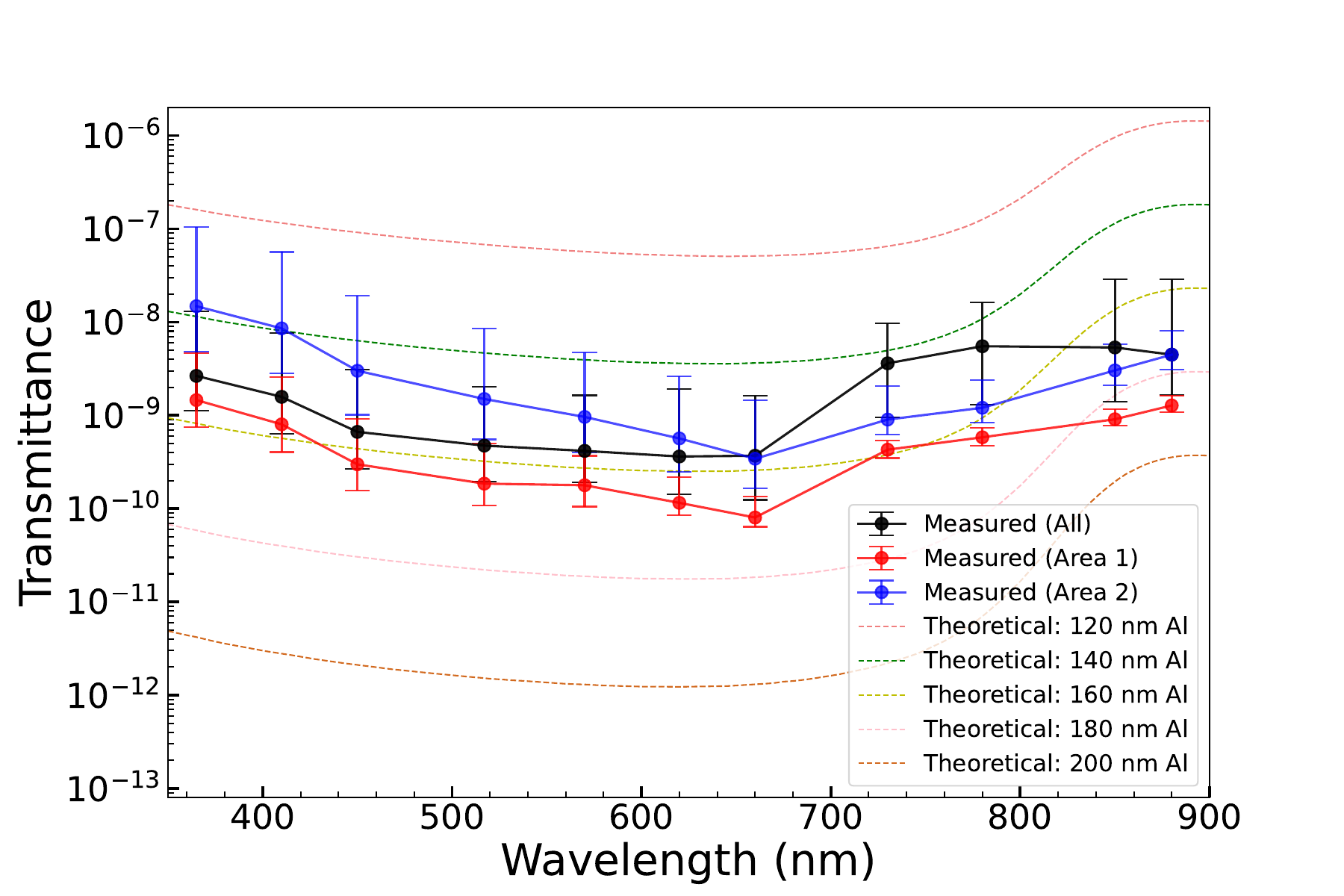}
\end{tabular}
\end{center}
\caption 
{ \label{fig:trans_vs_wl}
The measured transmittance of the coated aluminum layer at different wavelengths. The error bars represent the range between the 16\% percentile and the 84\% percentile of the transmittance among pixels. The dashed curves are the theoretical transmittance of a flat and uniform Al layer of different thicknesses, ranging from 120 nm to 200 nm.} 
\end{figure}

The transmittance of the coated aluminum layer was measured at different wavelengths, as shown in Fig.~\ref{fig:trans_vs_wl}. Data points are the median transmittance of pixels in the whole active area (black), in Area 1 (red), and in Area 2 (blue), respectively. The error bars represent the range between the 16\% percentile and the 84\% percentile of the transmittance among these pixels. The dashed curves in Fig.~\ref{fig:trans_vs_wl} are the theoretical results of Al layers of different thicknesses. In our theoretical calculation, we use the optical parameters of aluminum from \citet{McPeak_Al_optic_parameter} and follow the results of Section 14.4.1 in \citet{principles_of_optics}. We assume that the aluminum layer is flat and uniform and that light is incident vertically into the film in our calculation. The measurement results of the optical transmittance of the aluminum layer on the sCMOS sensor show that the effective thickness of this Al layer is around 140 to 160 nm, providing a transmittance of around $10^{-8}$ to $10^{-10}$ for optical lights from 365 to 880 nm.

The thickness of the aluminum layer is measured with a profilometer to be around 200 to 210 nm, which is consistent with the initial design of 200 nm. However, the effective thickness around 140 to 160 nm is much smaller than this value. The existence of pinholes and voids in the aluminum film resulting from the imperfect sputter process can significantly weaken the light shielding performance of the film. This is also true for free-standing films and Al-coated CCDs \citep{filters_review_Barbera_2022, eRosita_CCD_PhD_thesis}. Furthermore, a thin layer of Al on the surface of the film will be oxidized to $\rm{Al_{2}O_{3}}$ \citep{XMMNewton_EPIC_filter_free_15yearslater}, which is highly transparent to optical lights. Additionally, light leakage from the edges of the sensor can also increase the transmittance.

\section{Conclusions}
\label{sect:concls}
We customized a large-format sCMOS sensor, G\-1516\-BI, in 2019. To maximize its X-ray detection capabilities, a 200 nm thick aluminum layer was uniformly sputtered on the entrance window of this sensor to block optical light. This Al-coated sensor, named EP4K, was successfully fabricated in 2021. 

The basic performances of this Al-coated EP4K sensor were measured and compared to those of the uncoated G\-1516\-BI sensor. The readout noise of the EP4K sensor is around 2.5 $\rm{e^{-}}$ and the dark current is less than 0.01 $\rm{e^-/pixel/s}$ at $\rm{-30 ^{\circ}\!C}$. The ratio of single pixel events of the Al-coated sensor is $45.0\%$. The energy resolution can reach 153.2 eV at 4.51 keV and 174.2 eV at 5.90 keV at $\rm{-30 ^{\circ}\!C}$. These performances are similar to those of the uncoated sensor. This means that the properties and performances of this sCMOS sensor are not affected by the sputtered aluminum film on the sensor. 

The optical transmittance of the coated aluminum layer on the EP4K sensor is measured at different wavelengths. Despite the possible influence of light leakage, pinholes and voids, and aluminum oxidation, the effective thickness of this Al layer is around 140-160 nm, providing a transmittance of around $10^{-8}$ to $10^{-10}$ for optical lights from 365 to 880 nm, which is not only much lower than the requirement of $10^{-6}$ of the Wide-field X-ray Telescope (WXT) module on the EP satellite, but is also good enough for other space X-ray missions. 

The good X-ray performance and low optical transmittance of this Al-coated sCMOS sensor make it a promising choice for X-ray telescopes. The LEIA, which has been working in orbit for about one year, is equipped with four pieces of EP4K sensors \citep{LEIA_Zhang_APJL_2022, LEIA_RAA}. Furthermore, 48 pieces of EP4K sensors are used on the WXT on the EP satellite, which will be launched at the end of 2023 \citep{EP_2018, EP_2022}. In the future, to further demonstrate the suitability of the sensor for space applications, aging tests and several types of irradiation tests will be performed on this Al-coated sCMOS sensor.

\begin{acknowledgments}
This work is supported by the National Natural Science Foundation of China (grant No. 12173055) and the Chinese Academy of Sciences (grant Nos. XDA15310000, XDA15310100, XDA15310300, XDA15052100).
\end{acknowledgments}

%





\bibliography{library.bib}
\bibliographystyle{aasjournal}



\end{document}